\documentclass[a4paper,12pt]{article}
\usepackage{amsmath,amsfonts,amssymb,cite,graphicx}
\newcommand{\be}{\begin{equation}}
\newcommand{\ee}{\end{equation}}

\begin{document}

\begin{center}
{\Large\bf AdS/QCD models describing a finite number of excited
mesons with Regge spectrum}
\end{center}

\begin{center}
{\large S. S. Afonin\footnote{On leave of absence from V. A. Fock
Department of Theoretical Physics, St. Petersburg State
University, 1 ul. Ulyanovskaya, 198504, Russia. E-mail:
afonin24@mail.ru.}}
\end{center}

\begin{center}
{\it Institute for Theoretical Physics II, Ruhr University Bochum,
150 Universit\"{a}tsstrasse, 44780 Bochum, Germany}
\end{center}

\begin{abstract}
The typical AdS/QCD models deal with the large-$N_c$ limit of QCD,
as a consequence the meson spectrum consists of the infinite
number of states that is far from the real situation. Basing on
introduction of anharmonic corrections to the holographic
potential, the corrections whose existence has been recently
advocated, we construct a class of bottom-up holographic models
describing arbitrary finite number of states in the sector of
light mesons. Within the proposed approach, the spectrum of masses
square has the following properties: It is linear, $m_n^2\sim n$,
at not very large $n$, nonlinear at larger $n$, with the nonlinear
corrections being subleading in $1/N_c$, has a limiting mass, and
the number of states is proportional to $N_c$. The considered
holographic models reflect thereby the merging of resonances into
continuum and the breaking of gluon string at sufficiently large
quark-antiquark separation that causes the linear Regge
trajectories to bend down. We show that these models provide a
correct description for the spectrum of excited $\rho$-mesons.
\end{abstract}


\newpage

\section{Introduction}

Recently the holographic models of QCD have been successfully
applied to the description of non-perturbative physics of the
strong interactions~\cite{br1,son1,son2,pom}. For this reason it
is interesting to further investigate to what extent such a
phenomenological approach (often called AdS/QCD) is able to
describe the low energy QCD irrespective of whether it is related
to the original AdS/CFT correspondence~\cite{mald} or not.

Maldacena's hypothesis~\cite{mald} provided a promising way to
link the fundamental string theories in the low energy
approximation to the strongly coupled four-dimensional conformal
field theories living on the boundary of AdS space~\cite{wit}. As
the AdS/CFT method permits to obtain a theoretical control over
the strongly coupled gauge theories, the next natural step is to
extend the idea of holographic correspondence to the physical
gauge theories, such as QCD. This step was realized, albeit
speculatively, through the AdS/QCD models (see, {\it
e.g.},~\cite{forkel} for a brief review). The latter approach,
however, deals typically with the planar limit of QCD,
$N_c\rightarrow\infty$, {\it i.e.} it is rather restricted from
the very beginning (see discussions in~\cite{cohen} on that
point). As a consequence, the meson spectrum in the AdS/QCD models
consists of an infinite number of infinitely narrow states in a
sharp contradiction with what we observe experimentally: There are
a few of resonances in each channel, they have a finite width and
the discrete spectrum gradually merges into perturbative
continuum. In the program of relating the strings to the real
hadron physics, the next step is therefore called for --- the
development of holographic models describing a finite number of
hadrons merging into continuum. In the present Letter, we will
construct a class of bottom-up holographic models possessing this
property.

The paper is organized as follows. In Sect.~2, we formulate the
problem and specify the subject matter of this work. Sect.~3 is
devoted to the construction of a general design for the
holographic models describing a finite number of mesons in the
simplest case. Using the modern experimental data we show in
Sect.~4 that the found form of spectrum seems to be supported
experimentally. A simple particular realization of the proposed
idea is considered in Sect.~5. The results obtained, their
relation to other works, and the possible consequences are
discussed in Sect.~6. Sect.~7 concludes our analysis.

\section{Formulation of the problem}

The problem of primary importance in the non-perturbative QCD is
the description of spectrum of light hadrons. In this work, we
will be interested only in the spectroscopy of light resonances,
namely we will consider the light mesons because their spectrum is
well defined in the large-$N_c$ limit. Experimentally, the most
investigated sector is that of the $\rho$-mesons~\cite{pdg}, for
this reason we will restrict ourselves to the $\rho$-mesons only
since only in this case there is enough experimental data to make
more or less reliable fits.

The QCD coupling constant $\alpha_s$ is running, this fact makes
impossible to follow the principles of AdS/QCD correspondence
exactly, nevertheless, one can attempt to follow these principles
as close as possible. According to the averaged fit~\cite{pdg},
the value of $\alpha_s$ at the scale of the $\rho$-meson mass is
$\alpha_s(0.776~\text{GeV})\approx0.71$, at the scale of the first
radial excitation of the $\rho$-meson
$\alpha_s(1.465~\text{GeV})\approx0.35$, at the scale of the
highest observed radial excitations
$\alpha_s(2.300~\text{GeV})\approx0.28$. It is seen that in the
energy region where the most resonances reside the change of
$\alpha_s$ is moderate (about 20\%), on the other hand, the theory
is strongly coupled (there are resonances), hence, a weakly
coupled holographic model dual to QCD in that region has chances
to exist in some approximate sense. Below that region QCD is
strong but $\alpha_s$ changes rapidly (see, however, the recent
indications on the existence of infrared fixed point~\cite{deur}
suggesting $\frac{\alpha_s}{\pi}\lesssim1$), above the resonance
region the two-flavor QCD is approximately scale invariant  but is
in the weakly coupled regime, in both cases the principles of
AdS/QCD correspondence are badly violated. Thus, we are going to
describe holographically the resonance region that is restricted
by the infrared, $\Lambda_{\text{IR}}$, and the ultraviolet,
$\Lambda_{\text{UV}}$, cutoffs. The introduction of the UV-cutoff
separating the resonance region from the onset of the perturbative
continuum is crucial for our further analysis and distinguishes
our models from other AdS/QCD models existing in the literature.
The IR-cutoff is fixed by hand in the hard-wall
models~\cite{br1,son1,pol} in order to model the confinement, in
our case an effective IR-cutoff will be introduced in a different
way.

Considering QCD with $SU(N_c)$ gauge group, where $N_c$ is finite
and arbitrary, we must answer the following question: How many
resonances should we take into account? The question is not
trivial because, generally speaking, the cutoff
$\Lambda_{\text{UV}}$ depends on $N_c$. However, there is a simple
argument~\cite{aelast} that if the spectrum of masses square is
linear, $m_n^2\sim \mu^2 n$ (as expected from the
phenomenology~\cite{phen} and the semiclassical string models for
mesons~\cite{sv2}) the number of resonances should be proportional
to $N_c$. Indeed, in the Breit-Wigner parametrization of
resonances, the width is $\Gamma_nm_n$ and if this quantity
becomes equal to the distance between masses square of the
neighboring resonances, $\Gamma_nm_n\sim\mu^2$, we cannot speak
any more about resonances as they merge into continuum. On the
other hand, according to the semiclassical string (flux-tube)
models~\cite{sv2} $\Gamma_n \sim m_n/N_c$, hence, $\Gamma_nm_n\sim
m_n^2/N_c\sim \mu^2n/N_c$ and we obtain $n_{\text{max}}\sim N_c$.
This expectation will be an important input in what follows.

\section{Construction of model}

In order to advance in holographic description of real hadron
physics one usually extends the existing successful holographic
models. There are two complementary ways for such extensions,
either by introducing new operators/fields following the AdS/CFT
principles or by modification of the background metric. The former
seems to be more consistent but at the same time more difficult
and somewhat ambiguous at the present stage. The latter seems
rather {\it ad hoc} but simpler, for this reason we will exploit
this geometric way hoping that subsequently the results obtained
will be justified from a more fundamental point of view with the
help of introduction of appropriate operators/fields. We note in
advance that the first step in justifying our approach has been
already done in~\cite{afonin}.

The simplest 5d action describing the spectrum of vector mesons is~\cite{son1}
\begin{equation}
\label{1}
S=\int d^5x\sqrt{g}\left(-\frac{1}{4g_5^2}F_{MN}F^{MN}\right),
\end{equation}
where $g=|\det{g_{MN}}|$ and $F_{MN}=\partial_MV_N-\partial_NV_M$.
Generally speaking, from the 5d side we must have a Yang-Mills theory
with $SU(N_f)$ gauge group~\cite{son1} but for calculation of the mass spectrum
one retains the quadratic in fields part only so that the Abelian part
of $F_{MN}$ is enough for our purposes. The IR boundary condition is that
the action is finite at $z=\infty$. The metric is parametrized as
\begin{equation}
\label{2}
g_{MN}dx^Mdx^N=e^{2A(z)}(dz^2+\eta_{\mu\nu}dx^{\mu}dx^{\nu}),
\end{equation}
with $\eta_{\mu\nu}=\text{diag}(-1,1,1,1)$. The equation of motion for
action~\eqref{1} possesses a solution for the string modes $V_M(x,z)$
which is supposed to be dual to physical states of the gauge theory.
Fixing the gauge $V_z=0$, the corresponding equation reads~\cite{son2}
\begin{equation}
\label{3}
\partial_z(e^A\partial_zv_n)+q^2e^Av_n=0,
\end{equation}
where $v_n$ must be normalizable solutions for the 4d-transverse components
$V_{\mu}^T$ which exist only for discrete values of 4d-momentum $q^2=m_n^2$.
Performing the substitution
\be
v_n=e^{-A/2}\psi_n,
\ee
Eq.~\eqref{3} takes the form
of a Schr\"{o}dinger equation,
\begin{equation}
\label{4}
-\psi_n''+U(z)\psi=m_n^2\psi_n,
\end{equation}
with the potential
\begin{equation}
\label{5}
U(z)=\frac14(A')^2+\frac12A''.
\end{equation}

Thus the spectrum of the 4d gauge theory and the metric of the dual 5d theory
are related (up to boundary conditions) by the form of potential~\eqref{5}.
The fifth coordinate $z$ is known to correspond to the energy scale $Q$:
$Q\sim1/z$. Let us divide the potential $U(z)$ into the UV and IR parts,
\begin{equation}
\label{6}
U(z)=U_{\text{UV}}(z)+U_{\text{IR}}(z),
\end{equation}
where $U(z)\xrightarrow[z\rightarrow0]{}U_{\text{UV}}(z)$ and
$U(z)\xrightarrow[z\rightarrow\infty]{}U_{\text{IR}}(z)$.

As follows from the AdS/CFT correspondence, the vector wave functions must
have the UV asymptotics $\psi(z)\xrightarrow[z\rightarrow0]{}z^2$~\cite{pol},
this dictates $U_{\text{UV}}(z)\xrightarrow[z\rightarrow0]{}z^{-2}$. The usual
choice is $U_{\text{UV}}(z)=3/(4z^2)$ that yields $e^{2A(z)}=z^{-2}$. The
conformal isometry of the metric reflects then the conformal behavior of QCD
in the ultraviolet. The conformal invariance can be broken by two ways ---
either by introducing a hard cutoff $z_{\text{IR}}$~\cite{br1,son1,pol} (the resulting
spectrum $m_n\sim n$ does not agree with the phenomenology in this case) or
by introducing a nontrivial $U_{\text{IR}}(z)$. In order to obtain the desired
behavior $m_n^2\sim n$ one has to have $U_{\text{IR}}(z)\sim z^2$ at least at
$z\rightarrow\infty$, {\it i.e.} a potential of the linear oscillator type.
This idea was realized in the soft-wall models~\cite{son2} by means of introduction
of a dilaton field in action~\eqref{1}.

We want to have a spectrum of oscillator type at small $n$ and a
finite number of discrete energy levels. This is known to require
a certain anharmonicity at large $z$. The general form of the
spectrum given by such anharmonic potentials can be derived in a
model-independent way by analyzing the anharmonic corrections,
\begin{equation}
\label{7}
U_{\text{IR}}(z)=\omega^2z^2+\alpha z^3+\beta z^4,
\end{equation}
where we have taken into account the $\mathcal{O}(z^4)$ term
because the first anharmonic correction arising from the
$\mathcal{O}(z^3)$ term disappears~\cite{landau}. For the time
being we neglect the UV contribution $U_{\text{UV}}(z)$ (to be
discussed below).

Considering the anharmonic corrections as perturbations, Eq.~\eqref{4} leads to
the spectrum~\cite{landau}
\begin{equation}
\label{8}
m_n^2=2\omega(n+1/2)-\gamma(n+1/2)^2+\text{const},\qquad n=0,1,2,\dots
\end{equation}
where
\be
\gamma=\frac{3}{2\omega^2}\left(\frac{5}{2\omega^2}\alpha^2-\beta\right).
\ee
It can be seen straightforwardly that one has the spectrum of oscillator type at small
$n$ if $|\gamma|\ll\omega$ and a finite number of energy levels if $\gamma>0$. The
approximate independence of meson masses on $N_c$ imposes $\omega=\mathcal{O}(N_c^0)$.
In order to have the number of states proportional to $N_c$ we must have
$\gamma=\mathcal{O}(1/N_c)$, {\it i.e.}, the nonlinear corrections to the spectrum
are subleading in the large-$N_c$ counting.

Consider now the impact of UV region introducing
$U_{\text{UV}}(z)=\varepsilon/z^2$, $\varepsilon\rightarrow0$. The
boundary condition for the oscillator wave functions,
$\psi(\infty)=0$, is then supplied by $\psi(\delta)=0$,
$\delta\rightarrow0$. As a result the even levels in
spectrum~\eqref{8} disappear, one has $n=1,3,\dots$.
Replacing $n$ by
\be
k=n/2-1/2, \qquad k=0,1,2,\dots,
\ee
the leading in $N_c$ contribution is given by
\be
m_k^2\simeq4\omega(k+3/4)+\text{const}.
\ee
Enlarging $\varepsilon$ leads to a shift up of energy levels (see, {\it
e.g.}, the exact example for $\varepsilon=3/4$ in~\cite{son2}),
finally some of the highest levels can vanish. All this, however,
does not change the general conclusions about the qualitative
behavior of the spectrum.

We assume that the general form of potentials should be such that
there were no possibility for quantum tunneling to the deep
infrared, $z\rightarrow\infty$, which we do not know how to
interpret physically. Together with the requirement of finite
number of discrete energy levels this leads to the following
general form of our potentials: They look like a potential well
(with the right wall of the form $z^2$) at relatively small $z$
and gradually transform into a plateau at large $z$,
$U_{\text{IR}}(z)\xrightarrow[z\rightarrow\infty]{}\Lambda_{\text{UV}}^2$,
see Fig.~1.
The asymptotic behavior of the metric follows from Eq.~\eqref{5}:
\be
A(z)\xrightarrow[z\rightarrow\infty]{}\pm2\Lambda_{\text{UV}}z
\ee
(see Eq.~\eqref{2}).
\begin{figure}
\vspace{-1 cm}
\hspace{2 cm}
\includegraphics[scale=0.7]{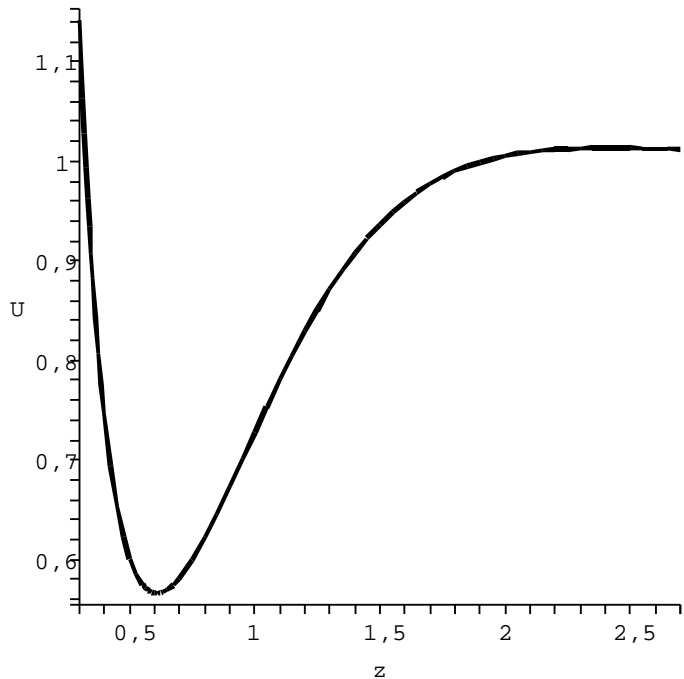}
\vspace{0 cm}
\caption{The general form of holographic potentials discussed in the present paper. The units are conditional.}
\vspace{0cm}
\end{figure}
In order to have the finite action we have to
choose the minus sign. The number of discrete energy levels is
determined by the value of $\Lambda_{\text{UV}}$ above which (plus
an arbitrary constant) one has a continuum and by the "width" of
the potential which we associate with the IR-cutoff
$\Lambda_{\text{IR}}$. As was said above, enlarging the factor in
front of $U_{\text{UV}}(z)$ one shifts up the energy levels, as a
result some of the highly excited states can disappear. The
quantity $\Lambda_{\text{IR}}$ determines (up to a presumably
small constant) an effective IR-cutoff $z_{\text{IR}}$ in the
coordinate space: In the vicinity of $z_{\text{IR}}$ the potential
well ceases to be of the form $z^2$ starting to transform into a
plateau.

\section{Comparison with the experiment}

Let us compare our suggestions with the experiment. In the Section
"Light Unflavored Mesons", the Particle Data~\cite{pdg} reports
five $\rho$-mesons: $\rho(770)$, $\rho(1450)$, $\rho(1700)$,
$\rho(1900)$, and $\rho(2150)$, the last two states are omitted
from the summary table but we include them into our fit as they
have been independently seen by several collaborations. Taking the
experimental masses of these resonances with reported errors we
can make the least square fit with the curve
\be
m_n^2=An^2+Bn+C
\ee
and estimate the errors in values of parameters, the result is (in
GeV$^2$)
\begin{equation}
\label{9}
m_n^2\approx(-0.09\pm0.02)n^2+(1.30\pm0.08)n+(0.71\pm0.02).
\end{equation}

In the Section "Other Light Unflavored Mesons", the Particle
Data~\cite{pdg} reports another two $\rho$-mesons: $\rho(2000)$
and $\rho(2270)$. The former was seen by one collaboration only,
for this reason we do not consider it. The latter was
independently seen by two collaborations, its mass is in a good
agreement with the fit~\eqref{9}.

Since the lowest three resonances are particularly well
established, it may be reasonable to display also the fit based on
these three states only,
\begin{equation}
\label{9b}
m_n^2\approx(-0.37\pm0.10)n^2+(1.91\pm0.18)n+0.06.
\end{equation}

Thus the nonlinear correction to the spectrum is in a qualitative agreement with our
analysis both in sign and in magnitude. Having fit~\eqref{9} one can obtain the
numerical values for the parameters of anharmonic potentials.

\section{A toy-model}

To estimate the typical values of input parameters let us consider the following
example of potential having the shape "potential well + plateau",
\begin{equation}
\label{10}
U(z)=\Lambda_{\text{UV}}^2\left(1-e^{-\Lambda_{\text{IR}}(z-z_0)}\right)^2+C,
\end{equation}
where $C$ is a constant. This is the Morse potential~\cite{morse} that has been widely used for
description of vibrations of nuclei (near $z_0$) in a diatomic molecule. In our case,
by shifting $z_0>0$ we can imitate $U_{\text{UV}}(z)$ in the region of applicability
of our models, thus the potential~\eqref{10} is a good interpolation for the class of
potentials described above. This potential has an advantage of being exactly solvable,
the corresponding spectrum is~\cite{morse}
\begin{equation}
\label{11}
m_n^2=2\Lambda_{\text{UV}}\Lambda_{\text{IR}}(n+1/2)-\Lambda_{\text{IR}}^2(n+1/2)^2+C,
\end{equation}
\begin{equation}
\label{12}
0\leq n\leq\frac{\Lambda_{\text{UV}}}{\Lambda_{\text{IR}}}-\frac12.
\end{equation}
Exploring fit~\eqref{9} we obtain the values of
parameters\footnote{As seen the numerical value of $C$ is small
and compatible with $C=0$ within the experimental errors. For
instance, as long as the mass of $\rho(1900)$ is not well
established, the value $C=0$ is achieved at
$m_{\rho(1900)}\approx1.93$~GeV, with the values of cutoffs being
affected slightly: $\Lambda_{\text{UV}}\approx2.32$~GeV,
$\Lambda_{\text{IR}}\approx0.31$~GeV.}:
$\Lambda_{\text{UV}}\approx2.34$~GeV,
$\Lambda_{\text{IR}}\approx0.30$~GeV, $C\approx0.04$~GeV$^2$. One
has then seven resonances with the masses (in GeV)
\be
m_n=\{0.84, 1.39,1.72, 1.96, 2.12, 2.24, 2.31\}.
\ee
According to our previous
discussions, the scaling of cutoffs in $N_c$ is
$\Lambda_{\text{UV}}=\mathcal{O}(\sqrt{N_c})$,
$\Lambda_{\text{IR}}=\mathcal{O}(1/\sqrt{N_c})$. It is interesting
to note that to the leading order in $N_c$ the mass of the ground
state is (neglecting a small $C$) the geometric mean of two
cutoffs.

Potential~\eqref{10} can be made a representative from our class
of potentials if we set $z_0=0$ and add
$U_{\text{UV}}(z)=\varepsilon/z^2$ but the model cannot be then
solved analytically. At small enough $\varepsilon$ and assuming
$\Lambda_{\text{UV}}\gg\Lambda_{\text{IR}}$ we obtain in this case
that the discrete levels reside in the potential well at
$z_{\text{UV}}<z<z_{\text{IR}}$, where
$z_{\text{UV}}\simeq\sqrt{\varepsilon}/\Lambda_{\text{UV}}$,
$z_{\text{IR}}\simeq1/\Lambda_{\text{IR}}$. At the choice
$\varepsilon=3/4$, the wave functions $v_n(z)$, where $n$ is small
enough ({\it i.e.} one has nearly linear spectrum) approximately
coincide with those found in~\cite{son2}.

\section{Discussions}

The proposed class of AdS/QCD models shares the attractive
features of both hard- and soft-wall holographic models: Like in
the hard-wall models, there is an (effective) IR-cutoff related to
confinement and there is no artificial dilaton the physical origin
of which is unclear, on the other hand, similarly to the soft-wall
models, one has approximately linear spectrum and the absence of
ambiguity in the choice of IR boundary conditions. However, if we
include the higher-spin fields in the way proposed in~\cite{son2},
the dilaton field seems to be inescapable\footnote{Assuming that
there are not additional fields in the action~\eqref{1} which are
dual to certain QCD operators.} if one wants to have the slope of
trajectories independent of spin $S$ and the relation
$m_{n,S}^2\sim n+S$ that fits well the known experimental
data~\cite{prc,sv}. The shape of the dilaton will be different in
our case, up to a factor in the exponent it will look like
$e^{-z}$ instead of $e^{-z^2}$ obtained in~\cite{son2}.

It should be noted that the coupling constant $g_5$ entering
action~\eqref{1} can be obtained by matching to the high-energy
asymptotics of QCD two-point correlators in the same way as
in~\cite{son1} (in the case of finite number of resonances, the
finite energy sum rules~\cite{fesr} should be used). We believe,
however, that such a matching is performed out the region of
applicability of the models under consideration. In any case, it
is not needed for the derivation of the mass spectrum we are
concerned.

An important result of the considered models is that the spectrum
condenses at high energies, {\it i.e.} the radial Regge
trajectories bend down and the discrete spectrum ends near the
point of zero slope. Within the flux-tube models of mesons, this
effect is usually interpreted as the string breaking, hence, the
proposed models are able to describe this effect: Up to
model-dependent constants, the string tension in our scheme is
(see Eq.~\eqref{11})
\be
\sigma\sim
\Lambda_{\text{UV}}\Lambda_{\text{IR}}(1-tn-t)=\mathcal{O}(N_c^0),
\ee
where
\be
t=\Lambda_{\text{IR}}/(2\Lambda_{\text{UV}})=\mathcal{O}(1/N_c),
\ee
{\it i.e.} $\sigma$ is a constant in the large-$N_c$ limit only.

The issue of non-linear corrections to the string-like spectrum
was discussed some time ago in the context of QCD sum rules in the
planar limit (see, {\it e.g.},~\cite{sv2,sum1,sum2}), in
particular such corrections were advocated to be exponentially
small, $e^{-bn}$ with $b>0$, in Ref.~\cite{sum1}. If one assumes
the scaling $b=\mathcal{O}(1/\sqrt{N_c})$ and expands the
exponential regarding $b$ as a small parameter, the form of
exponential correction will be compatible with our results.

In the minimal version, the considered class of models has 3
parameters: The factors in front of the harmonic and anharmonic
terms in the holographic potential (we can set $\alpha=0$ or
$\beta=0$ in Eq.~\eqref{7} without change of ensuing conclusions).
This is not very constraining as long as experimentally only a few
of states are known in each channel. We note, however, that the
first two parameters are expected to be universal for all
channels, hence, adding new channels in the model one introduces
only one new parameter for each tower of resonances --- a new
arbitrary constant --- or even less due to some
degeneracies~\cite{sv,prc}, thus such models will be much more
predictive if many kinds of mesons are described simultaneously.

Recently the necessity of anharmonic corrections to the
holographic potentials was emphasized in Ref.~\cite{afonin}
because they reflect holographically the contribution of the QCD
operators from the Operator Product Expansion for the correlators
of quark currents~\cite{svz}. It is interesting to make the
following observation: The study of the toy-model in Sect.~5 suggests
an intriguing possibility that the infinite series of anharmonic
corrections found in~\cite{afonin} might be summed up into some
simple function on the 5d side.

The introduction of UV-cutoff $\Lambda_{\text{UV}}$ may solve the
following problem in building the AdS/QCD models: The 5d action is
assumed to be local, {\it i.e.} the higher-derivative terms are
supposed to be suppressed. However, there is no understanding what
scale should suppress those terms. Within the AdS/QCD models put
forward in the present work, one can speculate that the
suppression is provided by powers of $1/\Lambda_{\text{UV}}$.

From the physical sense and numerical estimates for the IR-cutoff
$\Lambda_{\text{IR}}$ it is tempting to associate this quantity
with $\Lambda_{\text{QCD}}$ as in the hard-wall models~\cite{br1}.
According to a usual belief, however, $\Lambda_{\text{QCD}}$ is
nearly independent of $N_c$ and this provides an approximate
$m_n=\mathcal{O}(N_c^0)$ scaling for meson masses. In our case, the
situation is more tricky: Although
$\Lambda_{\text{IR}}=\mathcal{O}(1/\sqrt{N_c})$, the scaling
$m_n^2\sim\Lambda_{\text{IR}}\Lambda_{\text{UV}}=\mathcal{O}(N_c^0)$
holds due to the existence of the UV-cutoff
$\Lambda_{\text{UV}}=\mathcal{O}(\sqrt{N_c})$, thus we should
assume
\be
\Lambda_{\text{QCD}}^2\sim\Lambda_{\text{IR}}\Lambda_{\text{UV}}.
\ee
If the considered models indeed reflect the real QCD,
for the first time we are dealing with a situation when we can
learn something qualitatively new about QCD from the AdS/QCD
models. Specifically, our previous analysis suggests the following
mechanism for generating $\Lambda_{\text{QCD}}$
in the probable analytical solution for QCD: One introduces the
IR and UV cutoffs and takes the limit
$\Lambda_{\text{IR}}\rightarrow0$,
$\Lambda_{\text{UV}}\rightarrow\infty$ such that the product
$\Lambda_{\text{IR}}\Lambda_{\text{UV}}$ remains finite and should
be associated with $\Lambda_{\text{QCD}}^2$ that determines the mass
scale of the theory with massless quarks. Dealing with
an effective theory of QCD one works with finite
$\Lambda_{\text{UV}}$ and nonzero $\Lambda_{\text{IR}}$ that are
input parameters of the effective theory. Precisely this ideology
lies behind the models considered in the given work.

\section{Conclusions}

In the present paper, we have designed a class of AdS/QCD models
that describes a finite number of meson resonances whose spectrum
is approximately Regge-like. Assuming that the number of
distinguishable resonances depends linearly on $N_c$ we obtained
that the non-linear corrections to the Regge-like spectrum are
subleading in $1/N_c$. Within the given models, the non-linear
corrections stem from the anharmonic contributions to the
holographic potential that determines the mass spectrum. According
to the recent results~\cite{afonin} the anharmonic contributions
on the 5d side appear after taking into account the QCD operators
responsible for the masses of hadrons, therefore the design of the
considered models is somewhat justified from a more fundamental
standpoint. The presented approach and that of~\cite{afonin} shear
also the following important property: The slope of radial
trajectories is, up to a dimensionless factor of the order of
unity, a product of IR and UV cutoffs of the model, {\it i.e.} the
discrete mass spectrum is equally determined by both IR and UV
sectors of the theory. This interesting feature suggests that
something similar may happen in the exact solution for the mass
spectrum of real QCD.

\section*{Acknowledgments}

The work is supported by the Alexander von Humboldt Foundation. I
am grateful for the warm hospitality by Prof. Maxim Polyakov
extended to me at the Bochum University.

\end{document}